\documentclass[aps,preprint,pra,floatfix]{revtex4-1}
\usepackage{graphicx}
\usepackage{multirow}

\begin{document}
\title{Band gap tunning in BN-doped graphene systems with high mobility}
\author{T. P. Kaloni$^1$, R. P. Joshi$^2$, N. P. Adhikari$^2$, and U. Schwingenschl\"ogl$^2$}
\email{udo.schwingenschlogl@kaust.edu.sa}
\affiliation{$^1$KAUST, PSE Division, Thuwal 23955-6900, Kingdom of Saudi Arabia}
\affiliation{$^2$Central Department of Physics, Tribhuvan University, Kirtipur, Kathmandu, Nepal}

\begin{abstract}
Using density functional theory, we present a comparative study of the electronic properties
of BN-doped graphene monolayer, bilayer, trilayer, and multilayer systems. In addition,
we address a superlattice of pristine and BN-doped graphene. Five doping concentrations
between 12.5\% and 75\% are considered, for which we obtain band gaps from 0.02 eV
to 2.43 eV. We show that the effective mass varies between 0.007
and 0.209 free electron masses, resembling a high mobility of the charge carriers.
\end{abstract}

\maketitle
The outstanding electronic properties of graphene have caused huge interest in this material,
though the gapless features restrict to build graphene-based electronic devices, which could 
be alternative for existing silicon-based technology. Therefore, opening a finite and 
tunable band gap in graphene is demanded. Great efforts have been devoted to opening of 
a tunable gap in graphene. Various routes have been proposed, for example: superlattices 
of graphene with $h$-BN \cite{Zhou,kaloni,Giovannetti,Xu}, atomic layers of hybridized $h$-BN 
and graphene domains \cite{ajayan}, bilayer graphene \cite{park}, graphene nanowires 
\cite{son}, and twisted graphene \cite{li}. In gated bilayer graphene, a band gap of 250 meV 
and in graphene nanowires a band gap of 24 meV has been opened. The substrate induced band gap is found to be 52 
meV. In twisted graphene a band gap 270 meV have been reported. The many other ways also 
proposed to open a band gap in graphene, such as oxidation of graphene 
\cite{Bagri,Dai,Nourbakhsh}, hydrogenation \cite{Pumera}, and inducing of 
fluorine/boron-oxide in graphene \cite{Withers,udo2,udo3}. Recently, experimentally, a 
novel and promising way to open a band gap in graphene is proposed by controlling the 
domain size of in-plane heterostructures of graphene and $h$-BN \cite{ajayan1} 
   
The $h$-BN is wide band gap insulator with the same honeycomb structure as graphene 
\cite{Watanabe,Kubota}. It exhibits many  attractive properties such as high in-plane mechanical 
strength, excellent chemical inertness, and high thermal conductivity \cite{Balandin,Frank,Song,Kho,Chen}. 
It is suitable material to form the hybrid structure with graphene due to minimal lattice 
mismatch of about 1.6\% and both having same structural dimensionality. Hence, it is 
easy to achieve the hybrid structures of graphene and $h$-BN either in-plane manner 
or in stacking scheme as it has been demonstrated experimentally in 
Refs.\ \cite{ajayan,ajayan1}. Experimentally and theoretically the B/N doped graphene 
have been studied \cite{Sharifi,Han,Xue,sugata} and BN doped graphene leads to 
semiconducting \cite{Kim,Leenaerts,sugata1}. Still mechanism of the opening of the 
band gap in BN doped graphene, its electron effective mass, and mobility are unsettled.

Recently, the BN doped multilayer graphene and BN domain in graphene has been synthesized 
\cite{Kim,ajayan1}. Motivated by these experimental reports, 
we investigate the effect BN doping in the electronic properties of graphene monolayer, 
bilayer, trilayer, multilayer, and superlattice of pristine and BN-doped graphene, using 
density functional theory. The opening of the band gap found to be 0.02 eV to 2.43 eV for 
12.5\% to 75\% BN concentration. The calculated value of band gap are in good agreement 
with the available experimental and theoretical data. We estimate the electron effective 
mass for all the systems under study and found to be the value varies form 0.007 m$_e$ 
(which is smaller than that of reported for similar systems) to 0.209 m$_e$. We also 
estimate the mobility, which is as high as graphene on $h$-BN.

The results presented here are obtained from first-principles density functional theory 
in generalized gradient approximation \cite{Perdew}. 
The QUANTUM-ESPRESSO code is used with a plane-wave cutoff energy of 45 Ry \cite{paolo}. 
In addition, Van der Waals interaction has been taken into account via Grimmes scheme 
\cite{grimme}, which is expected to provide correct interlayer spacing and dispersion. 
A MonkhorstPack with a 8$\times$8$\times$1 k-mesh is used for all calculations under 
study. A 4$\times$4 supercell of graphene is used in our calculations. For all the 
structures, the atomic positions are fully relaxed until the energy convergence of 
10$^{-5}$ Ry and the force convergence of 10$^{-4}$ Bohr/Ry is achieved. 

The BN-doped graphene is classifies as (1) monolayer (see Fig.\ 1), (2) bilayer, (3) trilayer, 
(4) multilayer, and (5) superlattice. In all these systems, we have used five different BN-doping 
concentrations such as 12.5\% (2B2N), 25\% (4B4N), 43.75\% (7B7N), 56.25\% (9B9N), and 75\% (12B12N). 
In all of the mentioned doping concentration the ratio of B to N atom is same \cite{ajayan}. 
Turning to stacked structure: bilayer, trilayer, multilayer (superlattice of BN-doped graphene), 
and superlattice of pristine graphene with 
BN-doped graphene, there are two stacking schemes, such as AA and AB with different geometry. All of the possible 
configurations, N centered orientation in AB stacking is the most stable \cite{Giovannetti}. 
Hence, we only consider N centered AB stacking structures in our calculations. The 
distance between two layer in BN-doped bilayer and trilayer graphene is found to be 
3.34 \AA\, which is same as the separation between two layer in graphite. In all 
system under study, we found $a=b=9.94$ \AA. We use a vacuum layer of 12 \AA\ in 
order to avoid the interlayer interactions due to the periodic boundary conditions. 
Multilayer and superlattice of pristine and BN-doped graphene have orientation similar to that of 
BN-doped bilayer graphene in which the distance between two layer is  equal to 3.36 \AA\ and the length of 
the $c$ lattice to be 6.72 \AA. The modified C$-$C bond length for all the systems under study are 
vary from 1.40 \AA\ to 1.43 \AA, C$-$B bond length varies from 1.46 \AA\ to 1.49 \AA, 
C$-$N bond length varies from 1.36 \AA\ to 1.40 \AA, and B$-$N bond length varies from 
1.40 \AA to 1.45 \AA, while for pristine graphene C$-$C bond length is 1.42 \AA\ and 
B$-$N bond length is 1.44 \AA\ for pristine $h$-BN. The bond angle in BN-doped systems 
varies from 117$^\circ$ to 122$^\circ$, which is 120$^\circ$ for pristine graphene/$h$-BN.

\begin{figure}[ht]
\includegraphics[width=0.3\textwidth]{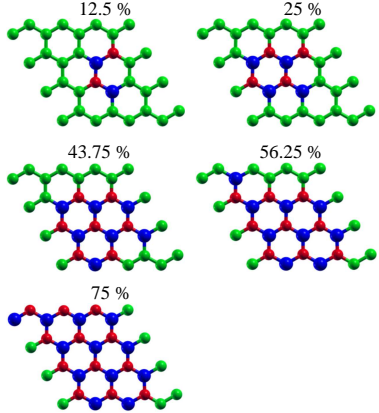}
\caption{Supercells used for modeling different BN concentrations.}
\end{figure}

To understand the stability of the structure and to compare the energetics we calculate 
the cohesive energy. The cohesive energy is calculated 
as $E_{coh}= \frac{E_{cell}-n\cdot E_C-m\cdot E_B-p\cdot E_N}{n+m+p}$, where E$_{coh}$, 
cohesive energy, E$_{cell}$ is the total energy of BN doped 4$\times$4 supercell, E$_C$, 
E$_B$, and $E_N$ are the total energy of isolated C, B, and N atoms, which is calculated 
by placing a single atom in a large and n, m, p are the total number of C, B, and N atoms,
respectively. The calculated values of cohesive energy for all the systems under study are 
presented in Table I. The cohesive energy for BN-doped monolayer ranges from 7.76 eV/atom 
to 8.02 eV/atom for 12.5\% to 75\%, which value is 7.75 eV/atom to 8.00 eV/atom for both 
bilayer and trilayer, and 7.75 eV/atom to 8.12 eV/atom for multilayer and superlattice of pristine and 
BN-doped graphene. These value slightly higher than that of the cohesive energy of 7.41 eV/atom for graphite 
\cite{Greenwood}, close to superlattices of graphene and $h$-BN \cite{kaloni}, and also 
well agree with the obtained value of the cohesive energy of $h$-BN \cite{ciraci}. From 
these obtained data for cohesive energy, we can conclude that all the system under study are highly feasible.   
 

The band gap is an intrinsic property of semiconductors, which indeed hugely determines
the transport and the optical properties. It should have a key role in modern device 
industry and technology. In this contrast, we focus the sensitivity of the electronic 
properties of the BN-doped thicker and thinner graphene by varying the BN concentration 
of 12.5\% to 75\%.  The calculated electronic structures of pristine graphene and 
BN-doped thinner and thicker graphene along the $\Gamma$-K-M-$\Gamma$ path is addressed 
in Fig.\ 2. The top row represents a monolayer graphene doped by 12.5\% to 75\% of BN 
atoms, second row represents bilayer, third row represents trilayer, fourth row 
represents multilayer, and fifth row represents superlattice of pristine graphene and BN-doped 
graphene with the same concentration as monolayer. We obtain the band gap of 0.42 to 2.43 eV for 12.5\% to 75\%,
 respectively, for all the systems under study, see Table I. The increasing of the band
 gap with respect to the BN concentration in graphene is addressed in Fig.\ 3. The
 opening of the band gap is attributed by the broken sublattice symmetry of the pristine
 graphene due the presence of the BN pair. Recently, the experimental setup for such
 level of doping is established by mass production of graphene samples using CH$_4$ gas
 as C source and ammonia borane precursor for BN source, which are introduced as a vapor
 in the whole duration of graphene growth \cite{Chang}. The concentration of the BN in
 graphene sample can be varied and controlled by controlling annealing temperature of
 the precursor. This process can be monitored by the X-ray photoelectron spectroscopy,
 which indeed confirms the monotonus increment in the BN concentration by the rising of 
the annealing temperature of the precursor. The authors of Ref.\ \cite{Chang} measured
 the band gap of about 0.6 eV for low concentration of the BN in the graphene sample,
 which in fact very close to our calculated band gap for the low concentration of 12.5\%.
 The another reason for the opening of the band gap is the change in the on-site energy of 
graphene due to the isoelectronic co-doping of BN pairs \cite{Liu1}. The obtained value of
 the band gap is increasing monotonically with increasing the BN pairs in the graphene 
sheet, due to the changing the on-site energy of the more and more carbon atoms in the system. 

The electronic band structure of undoped and the BN-doped bilayer and trilayer graphene by 12.5\% to 
75\% are shown in second and third rows of the Fig.\ 2. The pristine bilayer/trilayer graphene is 
metallic in nature \cite{Zhang1,Freitag}. The variation of the band gap 
follows the similar trend as in the case of monolayer but with slightly smaller gap, 
see Table I and Fig.\ 3. The band gap is decreasing with increasing the number of layers 
due the fact that the inversion symmetry broken for the bilayer/trilayer BN-doped graphene. 
The band gap opening in pristine bilayer graphene by the breaking of the inversion 
symmetry due to the application of the finite electric field has been demonstrated 
experimentally \cite{Zhang1}, a band gap of 250 meV has been achieved. However, in 
our calculations, by BN-doping in bilayer/trilayer graphene a variable band gap of 
0.04 eV to 2.20 eV is obtained, which can be enhanced and controlled by applying the 
electric field as it has been investigated for pristine bilayer graphene. The system 
we have studied here have been synthesized experimentally \cite{ajayan,ajayan1,Kim}. 

\begin{figure}[ht]
\includegraphics[width=0.7\textwidth]{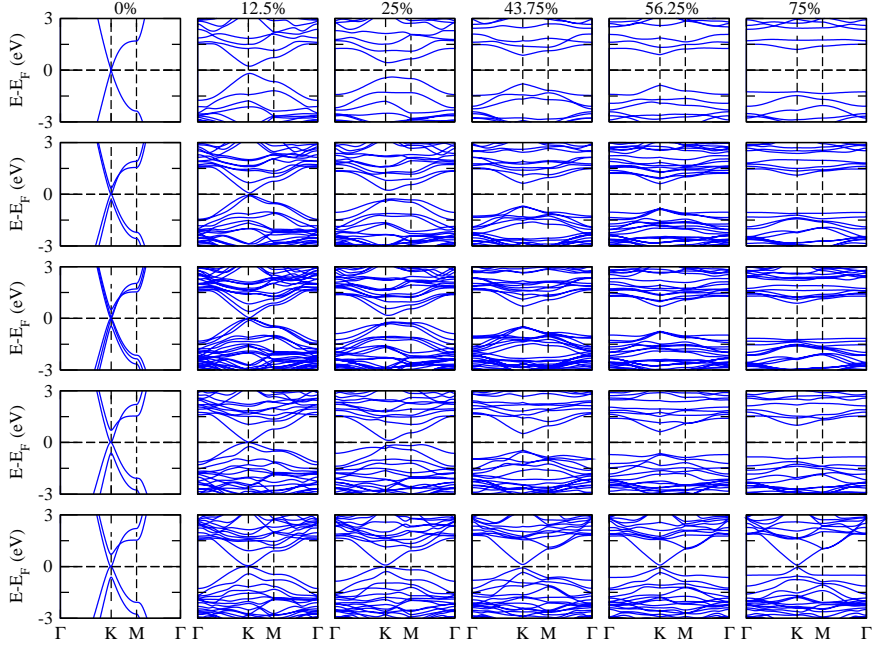}
\caption{Electronic band structure of BN-doped graphene for
concentrations between 0\% (red lines) and 75\% (blue lines). First row: monolayer, second row: bilayer,
third row: trilayer, fourth row: multilayer, fifth row: superlattice of pristine and
BN-doped layers. Note that the high symmetric points of the $4\times4$ supercell are folded back to the same points of the $1\times1$ cell. }
\end{figure}

The quantization of the energy states in the potential wells are important to understand
 the characteristic features of the superlattices. Which is always depends on the barrier 
height, when the barrier hight is small then 2 \AA\ the electronic energy bands should split
 into quasi-two-dimansional subbands 
like in conventional superlattices, for example InAs/GaSb. Moreover, if the barrier height
 is  greater than 2 \AA, the possibility of the overlapping of the neighbouring wavefunctions
 become suppressed. This give rise the the separate quantum wells of the superlattices
 \cite{Xu}. In our case the superlattices having a barrier height of greater than 2 \AA\ 
(i.e. 3.36 \AA). Our calculated band gap for superlattice is smaller than that of BN doped
 monolayer, bilayer, and trilayer graphene, see the electronic band structure addressed in
 fourth row of the Fig.\ 2 and Table I. Following the same trend as the other cases discussed
 above, the band gap increasing with the increasing the concentration of the BN atoms, which
 is well agree with the available theoretical prediction \cite{Lin} and experimental
 observation \cite{Kim}. Essentially, the electronic band structure of the superlattice 
is presented in fifth row of the Fig.\ 2 and obtained data are summarized in Table I. In
 this case, the minimum band gap of 0.08 eV to the maximum of 0.13 eV is observed. These 
obtained values are well agree with the available data for similar systems 
\cite{kaloni,Giovannetti,Xu}. The superlattice with all the BN doping concentration
 would be a promising candidate to build electronic device in a low voltage limit \cite{Zhang1}.
All the systems with low BN doping concentration are good for thermoelectric applications.

\begin{figure}[ht]
\includegraphics[width=0.45\textwidth]{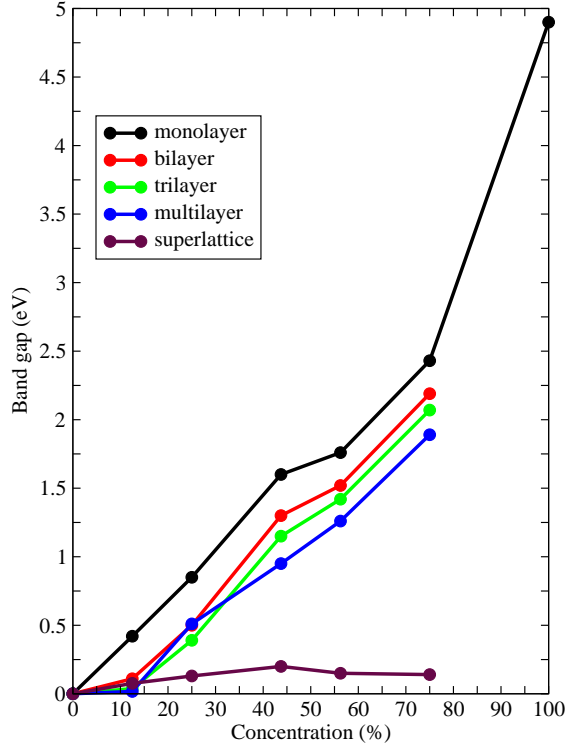}
\caption{Band gap as a function of doping for various BN-doped graphene systems.}
\end{figure}

From tight-binding approximation, the energy spectrum of gapped graphene, where the charge 
carriers behaves as massive Dirac particles is expressed as 
$E(k)=\pm v\sqrt{k_x^2+k_y^2+m^*{^2}v^2}$ \cite{Casolo}, where $\pm$ represents the
 conduction and valance bands, k$_x$ and k$_y$ are wave vector and m$^*$ is electron 
effective mass and hence band gap can be formulated as $E_g = 2m^*v^2$. Finally, we
 attempt to calculate the electron effective mass of the electron defined by 
$(m^{*})\sim\frac{E_g}{2v^2}$ \cite{apl}, where E$_g$ is the band gap and v is the Fermi
 velocity, which is found to be $\sim$0.8$\cdot10^{6}$ m/s for BN doped systems. It is
 reported that the interlayer spacing is the key point to control the band gap in bilayer
 graphene but with the increasing the band gap the value of m$^{*}$ is hugely increased,
 which is not good for device application because it suppress the carrier mobility. Hence,
 its very important to keep and control the m$^{*}$ as much as small. Our calculated electron
effective mass for all the systems under study is varies from 0.007 m$_e$ to 0.209 m$_e$,
 presented in Table I. The calculated m$^{*}$ really very small for low concentration limit
 of all the BN doped graphene systems. Interestingly the m$^{*}$ is lowest form all the
 concentration from 12.5\% to 75\% for superlattice of pristine and BN-doped graphene 
 because of smallest band gap 
opening in this system, and hence would be great candidate to built a electronic and 
opto-electronic devices, those can be operated in a low voltage/energy gap with a highly
 enhanced mobility \cite{magno}.

\begin{table}[ht]
\begin{tabular}{|c|c|c|c|c|c|c|}
\hline 
Quantity&Concentration (\%) & Monolayer & Bilayer  & Trilayer  & Multilayer  &  Superlattice \\
\hline 
\multicolumn{1}{|c|}{\multirow{5}{*}{E$_{coh}$ (eV)}}&12.5       & 8.02       & 8.00 & 8.00  & 8.12 &8.18	 \\ 
\cline{2-7} 
                                                  &25         & 7.94       & 7.93 & 7.93 & 7.92 & 8.15 \\ 
\cline{2-7} 
                                                  &43.75      & 7.85       &7.84& 7.84 & 7.83  & 8.11 \\ 
\cline{2-7} 
                                                  &56.25      & 7.78       & 7.79 & 7.77 & 7.77  &8.08 \\ 
\cline{2-7} 
                                                  &75         & 7.76       & 7.75 & 7.75 & 7.75 & 8.07 \\ 
\hline
\multicolumn{1}{|c|}{\multirow{5}{*}{E$_g$ (eV)}}  &12.5  & 0.42   & 0.11    & 0.04   &  0.02  & 0.08   \\ 

\cline{2-7}                                        &25          & 0.85   & 0.50    & 0.40   & 0.49   & 	0.12\\ 

\cline{2-7}                                        &43.75       & 1.46   &  1.30   & 1.15   & 0.96   & 0.20  \\ 
 
\cline{2-7}                                        &56.25  & 1.76   &  1.52   & 1.40   & 1.22   & 0.14  \\ 

\cline{2-7}                                        &75    & 2.43   & 2.20   & 2.08    &1.90    & 0.13  \\ 
\hline 
\multicolumn{1}{|c|}{\multirow{5}{*}{m$^*$ (m$_e$)}}&12.5 & 0.040 & 0.009   & 0.003 & 0.02   & 0.007   \\ 

\cline{2-7}                                        &25 & 0.073 & 0.043   & 0.034   & 0.042   &0.010   \\ 

\cline{2-7}                                        &43.75 & 0.125  &  0.111 & 0.098   & 0.082  & 0.017   \\ 

\cline{2-7}                                        &56.25 & 0.151  &  0.130   & 0.120  & 0.105   & 0.012   \\ 

\cline{2-7}                                        &75 & 0.209  & 0.189  & 0.178  & 0.163   & 0.011 \\ 
\hline
\multicolumn{1}{|c|}{\multirow{5}{*}{\begin{tabular}{c} $\mu$ \\ (m$^2$V$^{-1}$s$^{-1}$)\end{tabular}}}&12.5 & 0.84 & 3.71 & 6.68 & 1.67& 4.77   \\ 

\cline{2-7}                                        &25 & 0.46 & 0.78 & 0.48& 0.85  &3.34   \\ 

\cline{2-7}                                        &43.75 & 0.28 & 0.30& 0.34 & 0.40  & 1.97   \\ 

\cline{2-7}                                        &56.25 & 0.22& 0.26& 0.28 &0.32& 2.78   \\ 

\cline{2-7}                                        &75 & 0.16 &0.18  & 0.19 &0.20 & 3.04 \\ 
\hline
\end{tabular}
\caption{Cohesive energy, band gap, effective mass, and electron mobility
for various BN-doped graphene systems.}
\end{table}

We next calculate the electron mobility for all the systems under study. The electron 
mobility can be expressed as $\mu=\frac{e\tau}{m^*}$, where $\tau$ is the momentum 
relaxation time with a typical value of 1.9$\cdot10^{-13}$ s for Si doped graphene
 \cite{fedorov} and e is the charge of electron (1.6$\cdot10^-{19}$ c). The estimated
 $\mu$ varies from 0.18 m$^2$V$^{-1}$s$^{-1}$ to 6.68 m$^2$V$^{-1}$s$^{-1}$, see Table I.
 We found the $\mu^*$ is amazingly high for the BN-doped trilayer graphene which is 6
 times higher than that of SiO$_2$ gated pristine graphene \cite{Ishigami} and 1.11
 times higher than that of graphene on $h$-BN substrate \cite{hone}. We believe that
 our investigated systems should be excellent candidate for device application.

In conclusion, we performed the comparative study of electronic properties of BN doped 
monolayer graphene, bilayer, trilayer, multilayer, and superlattice of pristine and 
BN-doped graphene. Our results prevail that the band gap of graphene system can be tuned by 
BN-doping with different stacking pattern and concentrations. 
Due to the breaking of the sublattice symmetry by presence of B and N atoms a band gap 
of 0.02 eV to 2.43 eV is obtained for 12.5 \% to 75\%  BN-doping concentration. Our estimated electron 
effective mass for the systems under study is found to be varies form 0.007 m$_e$ to 
0.209 m$_e$, which is smaller than that of obtained for similar systems before. A giant 
mobility is estimated as compared to experimentally observed value for graphene on 
$h$-BN \cite{hone}, would have great impact on devices. Finally, our predict is to open 
finite and tunable band gap in BN-doped graphene with a giant mobility, which is 
promising for future graphene based electronics \cite{ajayan,ajayan1,Kim}.


\begin{thebibliography}{50}
\bibitem {Zhou} S. Y. Zhou, G.-H. Gweon, A. V. Fedorov, P. N. First, W. A. De Heer, D. -H. Lee, F. Guinea, A. H. Castro Neto, and A. Lanzara, Nat. Mater. \textbf{6}, 770 (2007).

\bibitem{kaloni} T. P. Kaloni, Y. C. Cheng, and U. Schwingenschl\"ogl, J. Mater. Chem. \textbf{22}, 919 (2012).

\bibitem{Giovannetti}G. Giovannetti, P. A. Khomyakov, G. Brocks, P. J. Kelly, and J. van den Brink, Phys. Rev. B {\bf 76}, 073103 (2007).

\bibitem{Xu} Y. Xu, Y. Liu, H. Chen, X. Lin, S. Lin, B. Yu, and J. Luo, J. Mater. Chem. {\bf 22}, 23821 (2012).

\bibitem{ajayan} L. Ci, L. Song, C. Jin, D. Jariwala, D. Wu, Y. Li, A. Srivastava, Z. F. Wang, K. Storr, L. Balicas, F. Liu, and P. M. Ajayan, 
Nat. Mater. {\bf 9}, 430 (2010).

\bibitem{park} C.-H. Park and S. G. Louie, Nano Lett. {\bf 10}, 426 (2010).

\bibitem{son} Y. W. Son, M. L. Cohen, and S. G. Louie, Phys. Rev. Lett. {\bf 97}, 216803 (2010).  

\bibitem{li} G. Li, A. Luican, J. M. B. Lopes dos Santos, A. H. Castro Neto, A. Reina, J. Kong, and E. Y. Andrei, Nat. Phys. {\bf 6}, 109 (2010). 

\bibitem{Bagri}A. Bagri, C. Mattevi, M. Acik, Y. J. Chabal, M. Chhowalla, and V. B. Shenoy, Nat. Chem. {\bf 2}, 581 (2010).

\bibitem{Dai}J. Dai and J. Yuan, Phys. Rev. B {\bf 81}, 165414 (2010).

\bibitem{Nourbakhsh}A. Nourbakhsh, M. Cantoro, T. Vosch, G. Pourtois, F. Clemente, M. H. van der Veen, J. Hofkens, M. M. Heyns, S. D. Gendt, 
and B. F. Sels, Nanotechnology {\bf 21}, 435203 (2010). 

    
 
\bibitem{Pumera}M. Pumera and C. H. A. Wong, Chem. Soc. Rev. {\bf 42}, 5987 (2013).

\bibitem{Withers}F. Withers, T. H. Bointon, M. Dubois, S. Russo, and M. F. Craciun, Nano Lett. {\bf 11}, 3912 (2011).

\bibitem{udo2}T. P. Kaloni, Y. C. Cheng, and U. Schwingenschl\"ogl, EPL {\bf 100}, 37003 (2013).

\bibitem{udo3} T. P. Kaloni, M. Upadhyay-Kahaly, R. Faccio, and U. Schwingenschl\"ogl, Carbon {\bf 64}, 281 (2013).

\bibitem{ajayan1} Z. Liu, L. Ma, G. Shi, W. Zhou, Y. Gong, S. Lei, X. Yang, J. Zhang, J. Yu, K. P. Hackenberg, A. Babakhani, J.-C. Idrobo, R. Vajtai, 
J. Lou, and P. M. Ajayan, Nat. Nanotechnol. {\bf 8}, 119 (2013).

\bibitem{Watanabe}  K. Watanabe,  T. Taniguchi, and  H. Kanda, Nat. Mater. \textbf{3}, 404 (2004).

\bibitem{Kubota} Y. Kubota,  K. Watanabe,  O. Tsuda, and  T. Taniguchi, Science \textbf{317}, 932 (2007).

\bibitem{Balandin} A. A. Balandin,  S. Ghosh, W. Bao, I. Calizo, D. Teweldebrhan, F. Miao, and C. N. Lau, Nano. Lett. \textbf{8}, 902 (2008).

\bibitem{Frank}  I. W. Frank, D. M. Tanenbaum, A. M. Van der Zande, and P. L. McEuen, J. Vac. Sci. Technol. \textbf{25}, 2558 (2007).

\bibitem{Song} L. Song, L. Ci, H. Lu, P. B. Sorokin, C. Jin, J. Ni, A.G. Kvashnin, D. G. Kvashnin, J. Lou, B. I. Yakobson, and  P. M. Ajayan, Nano. Lett. \textbf{10}, 3209 (2010).

\bibitem{Kho}  J. G. Kho, K. T. Moon, J. H. Kim, and D. P. J. Kim, Am. Ceram. Soc. \textbf{83}, 2681 (2000).

\bibitem{Chen}  Y. Chen, J. Zou, S. J. Campbell, and G. Le Caer, Appl. Phys. Lett. \textbf{84}, 2430 (2004).

\bibitem{Han} J. Han, L. L. Zhang, S. Lee, J. Oh, K.-S. Lee, J. R. Potts, J. Ji, X. Zhao, R. S. Ruoff, and S. Park, ACS Nano {\bf 7}, 19 (2013).

\bibitem{Sharifi} T. Sharifi, E. Gracia-Espino,	H. R. Barzegar,	X. Jia, F. Nitze, G. Hu, P. Nordblad, C.-W. Tai, and T. W\'agberg, 
Nat. Commun. {\bf 4}, 2319 (2013).

\bibitem{Xue}Y. Xue, D. Yu, L. Dai, R. Wang, D. Li, A. Roy, F. Lu, H. Chen, Y. Liu, and J. Qu, Phys. Chem. Chem. Phys. {\bf 15}, 12220 (2013).  

\bibitem{sugata} T. P. Kaloni and S. Mukherjee, Mod. Phys. Lett. B {\bf 25}, 1855 (2011).

\bibitem{Leenaerts}O. Leenaerts, H. Sahin, B. Partoens, and F. M. Peeters, Phys. Rev. B {\bf 88}, 035434 (2013).

\bibitem{sugata1} T. P. Kaloni and S. Mukhergie, J. Nanopart. Res. \textbf{14}, 1059 (2012)

\bibitem{Kim} S. M. Kim, A. Hsu, P. T. Araujo, Y.-H. Lee, T. Palacios, M. Dresselhaus, J.-C. Idrobo, K. K. Kim, and J. Kong, 
Nano Lett. \textbf{13}, 933 (2013).

\bibitem{Perdew} J. P. Perdew, K. Burke, and M. Ernzerhof, Phys. Rev. Lett. \textbf{77}, 3865 (1986).

\bibitem{paolo}P. Giannozzi, S. Baroni, N. Bonini, M. Calandra, R. Car, C. Cavazzoni, D. Ceresoli, G. L. Chiaro
tti, M. Cococcioni, I. Dabo, A. Dal Corso, S. de Gironcoli, S. Fabris, G. Fratesi, R. Gebauer, U. Gerstmann, C. 
Gougoussis, A. Kokalj, M. Lazzeri, L. Martin-Samos, N. Marzari, F. Mauri, R. Mazzarello, S. Paolini, A. Pasquare
llo, L. Paulatto, C. Sbraccia, S. Scandolo, G. Sclauzero, A. P. Seitsonen, A. Smogunov, P. Umari, and 
R. M. Wentzcovitch, J. Phys. Condens. Matt. {\bf 21}, 395502 (2009).  

\bibitem {grimme} S. Grimme, J. Comput. Chem. \textbf{27}, 1787 (2006).

\bibitem{Greenwood} N. N. Greenwood and A. Earnshaw, Chemistry of the Elements (Pergamon, Oxford, 1984).

\bibitem{ciraci} M. Topsakal, E. Akt\"urk, and S. Ciraci, Phys. Rev. B {\bf 79}, 115442 (2009).

\bibitem{Chang} C. Chang, S. Kataria, C.-C. Kuo, A. Ganguly, B.-Y. Wang, J.-Y. Hwang, K.-J. Huang, W.-H. Yang, S.-B. Wang, C.-H. Chuang, 
M. Chen, C.-I. Huang, W.-F. Pong, K.-J. Song, S.-J. Chang, J.-H. Guo, Y. Tai, M. Tsujimoto, S. Isoda, C.-W. Chen, L.-C. Chen, and K.-H. Chen, 
ACS Nano {\bf 7}, 1333 (2013).

\bibitem{Liu1}L. Liu and Z. Shen, Appl. Phys. Lett. {\bf 95}, 252104 (2009).

\bibitem{Freitag} M. Freitag, Nat. Phys. {\bf 7}, 596 (2011).

\bibitem{Zhang1}Y. Zhang, T.-T. Tang, C. Girit, Z. Hao, M. C. Martin, A. Zettl, M. F. Crommie, Y. R. Shen, and F. Wang, Nature {\bf 459}, 820 (2009).

\bibitem{Lin}B. Xu, Y. H. Lu, Y. P. Feng, and J. Y. Lin, J. Appl. Phys. {\bf 108}, 073711 (2010).

\bibitem{Casolo}S. Casolo, R. Martinazzo, and G. F. Tantardini, J. Phys. Chem. C {\bf 115}, 3250 (2011).

\bibitem{apl} Y. Fan, M. Zhaoa, Z. Wang, X. Zhang, and H Zhang, Appl. Phys. Lett. {\bf 98}, 083103 (2011).

\bibitem{magno} R. Magno, E. R. Glaser, B. P. Tinkham, J. G. Champlain, J. B. Boos, M. G. Ancona, and P. M. Campbell, 
J. Vac. Sci. Technol. B {\bf 24}, 1622 (2006).

\bibitem{fedorov} D. V. Fedorov, M. Gradhand, S. Ostanin, I. V. Maznichenko, A. Ernst, J. Fabian, and I. Mertig, 
Phys. Rev. Lett. {\bf 110}, 156602 (2013).

\bibitem{Ishigami}J.-H. Chen, C. Jang, M. Ishigami, S. Xiao, W. G. Cullen, E. D. Williamsa, M. S. Fuhrera, Solid State Commun. {\bf 149}, 1080 (2009).

\bibitem{hone}C. R. Dean, A. F. Young, I. Meric, C. Lee, L. Wang, S. Sorgenfrei, K. Watanabe, T. Taniguchi, P. Kim, K. L. Shepard, and J. Hone, Nat. Nanotechnol. {\bf 5}, 722 (2010). 
\end{thebibliography}
\end{document}